\documentclass[aps,prl,twocolumn,10pt,superscriptaddress]{revtex4-1}
\usepackage[pdftex]{color,graphicx}

\begin{document}

\title{Electronic properties of two-dimensional rectangular graphyne based on phenyl-like building blocks}

\author{Anderson Gomes Vieira}
\affiliation{Departamento de F\'isica, Universidade Federal do Piau\'i, CEP 64049-550, Teresina, Piau\'i, Brazil} 
\author{Marcelo Lopes Pereira Júnior}
\affiliation{Department of Electrical Engineering, Faculty of Technology, University of Brasília, Brasília, Brazil} 
\author{Vincent Meunier}
\affiliation{Department of Engineering Science and Mechanics, Pennsylvania State University, State College, PA, USA}
\author{Eduardo Costa Gir\~ao}
\affiliation{Departamento de F\'isica, Universidade Federal do Piau\'i, CEP 64049-550, Teresina, Piau\'i, Brazil}

 \date{\today}
\begin{abstract}
A rectangular graphyne sheet is composed of units similar to phenyl rings that are linked by acetylenic chains, as in hexagonal $\gamma$-graphyne. This system is organized over a rectangular lattice similar to that of the recently synthesized biphenylene network. We investigate the stability of this sheet from different perspectives and study its electronic structure. Rectangular graphyne is a semiconducting system in its pristine form and features a pair of highly localized states. These characteristics are correlated with the structural anisotropy of the system, since its frontier states behave like quasi--1D states embedded in the 2D lattice. We further consider modified systems in which longer acetylenic links are introduced. We discuss how a strategic choice of the position of these longer bridges can lead to specific changes of the electronic structure of the rectangular graphyne sheet. 
\end{abstract}

\maketitle

\section{Introduction}
The science and engineering of nanostructured materials have become central in the development of new platforms for future electronic applications. Among those materials, nanocarbons occupy a special position, as they can be found in all three low-dimensional configurations, ranging from 0D (\textit{e.g.}, fullerenes), 1D (\textit{e.g.}, nanotubes, and nanoribbons), to 2D (\textit{e.g.}, graphene)~\cite{meunier2016}. In two dimensions, graphene is the most striking example of a test bed for new fundamental physics and potential for applications~\cite{noorden2011,kumar2022,urade2023,meunier2022}. It features an electronic structure with Dirac cones~\cite{castroneto2009} and hosts a variety of interesting phenomena such as Klein tunneling~\cite{zhang2022} and high carrier mobility~\cite{wang2013}. The need to open a band gap in graphene and tune its electronic properties has motivated many studies related to structural modifications, including variations in nanoribbon structures~\cite{song2019}, the introduction of defects~\cite{terrones2012}, and the formation of other carbon heterostructures with different chemical species~\cite{cai2014}.

Another pathway to expand the collection of known properties of carbon at the nanoscale is the proposal of new two-dimensional allotropes with structural details markedly different from those of graphene. The past two decades have witnessed the proposal of several new carbon lattices with a broad set of electronic behaviors~\cite{crespi1996,terrones2000,hudspeth2010,wang2015,dossantos2023}. Although most of these systems have only been theoretically proposed, the synthesis of 2D biphenylene (BPN)~\cite{fan2021} and other nanoribbons with non-hexagonal rings~\cite{fan2019} has highlighted their relevance in the field of nanoscience. Many of these systems are solely composed of $sp^2$ atoms, but exploring other hybridized states of carbon has further expanded the family of 2D nanocarbons as well. Graphyne (GY) systems are among the most representative examples in this regard~\cite{ivanovskii2013,li2020}. In addition to being considered in many theoretical proposals, recent advances in graphyne science include the experimental realization of some representative systems~\cite{matsuoka2019,rabia2020,hu2022}. Graphyne systems consist of a carbon layer containing a mixture of $sp$ and $sp^2$ carbon atoms. These lattices are usually derived in some way from the structure of graphene, by introducing acetylenic units ($-C\equiv C-$) in lieu of selected $C-C$ bonds from the honeycomb lattice~\cite{malko2012,wu2013}. The study of graphynes has expanded to lattices different from graphene, such as systems with tetragonal symmetry~\cite{yin2013}, layers including pentagonal and heptagonal pores~\cite{oliveira2022}, among others~\cite{nulakani2017,tromer2023mechanical}.

The $\gamma$-GY monolayer is an interesting case structure in the broader context of efforts to open a band gap in graphene-like structures. Simulations show that it is a semiconducting system~\cite{kang2011} and it has recently been synthesized~\cite{hu2022}. The structure of a $\gamma$-GY layer is composed of hexagonal rings of $sp^2$ hybridized atoms linked to each other by $-C\equiv C-$ bridges in a hexagonal lattice. Similarly to how the $\gamma$-GY system can be thought of as a structural variation of a graphene layer, we propose a two-dimensional graphyne system that can be viewed as a variation of BPN, which is an already synthesized $sp^2$ structure~\cite{fan2021}. In the proposed system (which we will call r$\gamma$GY), hexagonal $sp^2$ rings are linked to each other by acetylenic links in a manner similar to the linking hierarchy of another synthesized structure, the hexagonal porous graphyne (HGY)~\cite{liu2022}. We use theoretical calculations to demonstrate that r$\gamma$GY is a stable system and a semiconducting material with highly anisotropic electronic properties. We show that the electronic structure of the system can be tuned with the introduction of longer acetylenic chains.

\section{Methods}

We use Density Functional Theory (DFT)~\cite{hohenberg1964,kohn1965}, as implemented in the SIESTA code~\cite{Soler2002} to compute the physical properties of r$\gamma$GY. The generalized gradient approximation (GGA) is adopted for the exchange-correlation functional according to the Perdew-Burke-Ernzerhof (PBE) parameterization~\cite{perdew1996}. Core electrons are represented using norm-conserving Troullier-Martins pseudopotentials~\cite{troullier1991}, and valence electrons are described in terms of a double-$\zeta$ polarized (DZP) basis set. The grid for real-space integration is defined by a 400 Ry mesh cutoff. A vacuum region of 20~\AA~is included along the direction perpendicular to the molecular planes of the two-dimensional systems to avoid interactions between periodic images. A total of $30\times32$ Monkhorst-Pack $k$ points are used to perform Brillouin zone (BZ) integrations in the r$\gamma$GY system. Other systems studied are sampled with $k$ grids of similar density. Structural optimizations are performed without constraints, considering a maximum threshold force of 0.01~eV/\AA~for each atom and a maximum tolerance of the stress component of 0.1 GPa for the optimization of the lattice parameters.

To evaluate the dynamical stability of r$\gamma$GY, we computed its phonon band structure. For this part of the study, the calculations were performed with DFT-based simulations using the GPAW code~\cite{mortensen2005,enkovaara2010}. The force constants were obtained by the finite difference method (with a displacement distance of 0.01~\AA) together with the PHONOPY package~\cite{togo2015}. GPAW uses the grid-based projector augmented wave (PAW) method in the description of electron-ion interactions~\cite{blochl1994}, while the manipulation and analysis of atomic simulations were performed with the aid of the atomic simulation environment (ASE) package~\cite{larsen2017}. We also used the GGA-PBE functional for the GPAW simulations. The plane-wave basis set was defined according to a 500~eV energy cutoff, and a smearing of 0.1 eV was used to compute Fermi-Dirac occupation. We used a $2\times2\times1$ Monkhorst–Pack k-point grid to sample the Brillouin-zone of the $3\times3\times1$ supercell. Atomic coordinate optimization was carried out with a quasi-Newton method according to the maximum threshold values of 1~meV/\AA~for atomic forces and 0.01~GPa for stress, with the structures further symmetrized with PHONOPY.

\section{Results and Discussion}
\begin{figure}[ht!]
\includegraphics[width=\columnwidth]{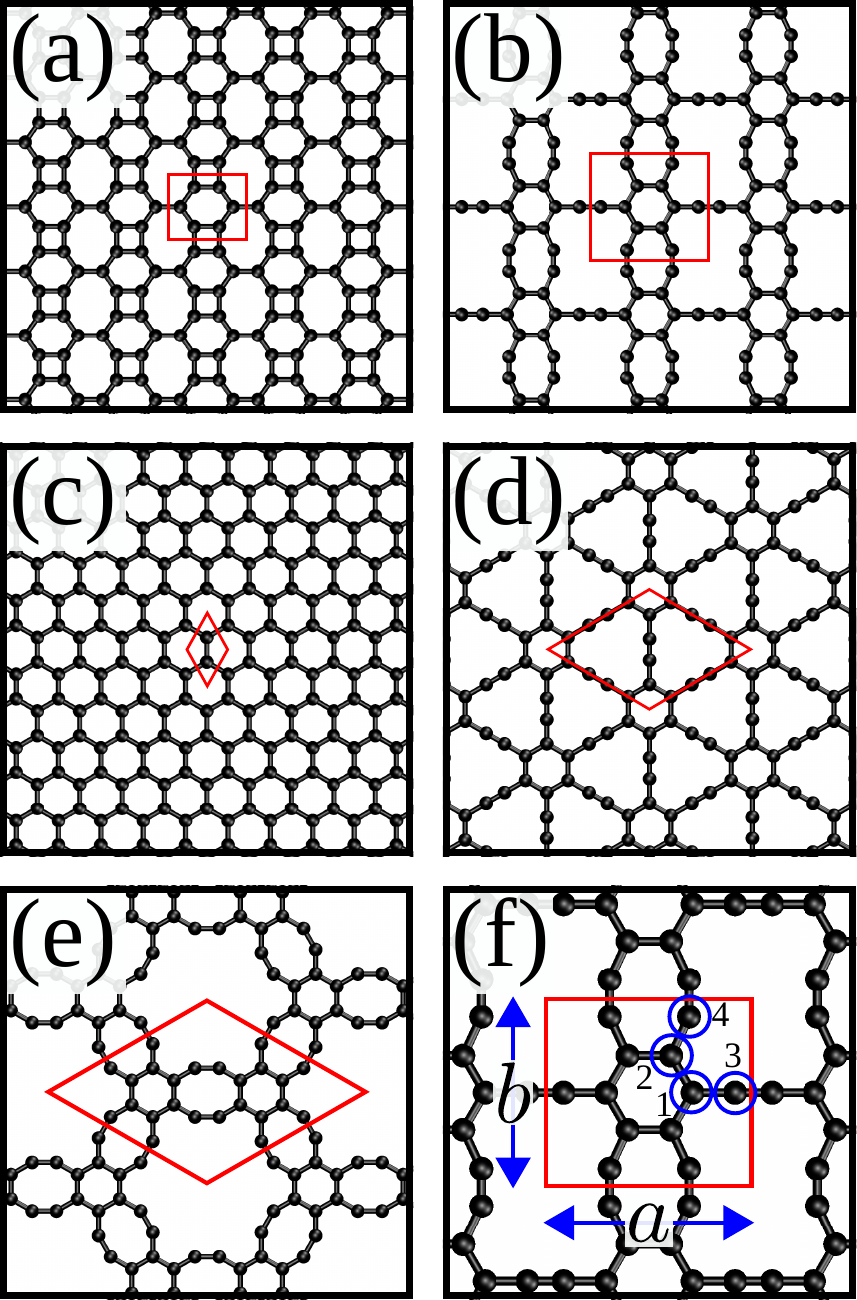}
\caption{Ball-and-sticks representation of the atomic structure of (a) 2D biphenylene (BPN), (b) r$\gamma$GY, (c) graphene, (d) $\gamma$-graphyne, and (e) hexagonal porous graphyne (HGY). In each case, the primitive unit cell is highlighted by a red rhombus or polygon. In (f), we highlight details of the atomic structure of r$\gamma$GY, namely its lattice parameters $a$ and $b$, its unit cell (red rectangle), and the identification of the non-equivalent atoms in the structure (labeled 1, 2, 3, and 4).}
\label{fig-01}
\end{figure}

The atomic structure of the GY sheet studied here is illustrated in Fig.~\ref{fig-01}b. Two different perspectives were considered as the basis for the proposal of the structure of interest. First, the system can be viewed as theoretically originating from a BPN sheet (illustrated in Fig.~\ref{fig-01}a) where some $C-C$ bonds are substituted by acetylenic bridges while preserving the hexagonal rings formed solely by $sp^2$-hybridized atoms. This is analogous to the way a $\gamma$-GY sheet (depicted in Fig.~\ref{fig-01}d) can be generated from a graphene layer (Fig.~\ref{fig-01}c). Taking into account this hypothetical construction, we call the structure shown in Fig.~\ref{fig-01}b ``$\gamma$-BPN-GY''. Alternatively, its assembly can be thought of as similar to that derived from holey graphyne (HGY, illustrated in Fig.~\ref{fig-01}e), where hexagonal rings are linked by pairs of acetylenic bridges in a hexagonal network. However, in the system studied here, this assembly with pairs of $-C\equiv C-$ links is associated with a rectangular network, resulting in a rectangular version of HGY. Using a recently proposed standard taxonomy~\cite{girao2023}, this system can be labeled as $r4^1_26^1_08^1_4$-graphyne. Hereafter we will simply refer to the studied structure as r$\gamma$GY, with ``r'' standing for its rectangular lattice and $\gamma$ referring to the fact that it contains a set of hexagonal rings made fully of $sp^2$ atoms, as in $\gamma$-GY. We note that the r$\gamma$GY pores originating from the BPN tetragons have a total of 8 atoms each (4 $sp$ and 4 $sp^2$ atoms). However, to simplify the terminology used in the discussion of the results, we will conventionally refer to the $C-C\equiv C-C$ sectors of this r$\gamma$GY pore as a single edge and refer to these pores as 4-membered or simply tetragonal units. Similarly, the r$\gamma$GY pores originating from BPN's octagons will be referred to as 8-membered or simply octagonal pores (even though they contain a total of 16 carbon atoms each). 

The structure's lattice parameters are illustrated in Fig.~\ref{fig-01}f and their values are $a=6.92$~\AA~and $b=6.28$~\AA. Its symmetries are contained in the $pmm2$ wallpaper group, which includes a set of $C_2$ symmetry axes at the 1) center, 2) corners, and 3) mid-points of the edges of the rectangular unit cell represented in Fig.~\ref{fig-01}f. A pair of orthogonal mirror planes pass at each of these $C_2$ axes. In Fig.~\ref{fig-01}f, we also identify the four nonequivalent atoms in the structure (blue circles labeled as 1, 2, 3, and 4). The bonds forming the hexagons ($d_{12}$ and $d_{22}$) are $\sim1.42$~\AA~and $\sim1.44$~\AA~long, respectively. These bond lengths are very similar to each other and close to the carbon-carbon distance in graphene. The hexagonal rings in r$\gamma$GY show much closer similarity to the hexagonal rings from graphene than when we compare graphene and BPN, as the carbon-carbon bond lengths in the BPN hexagons are $\sim1.41$~\AA~and $\sim1.46$~\AA, as calculated by the same methods. In fact, it has been discussed that non-hexagonal rings are better accommodated in graphyne sheets than in full-$sp^2$ layers with similar geometry, as the acetylenic bridges and the large pore structure of graphynic systems allow better opportunities for the release of strain over the structure~\cite{oliveira2022}. This is the case for the 4- and 8-membered rings from BPN. In BPN we observe bond angles of $90^\circ$ (inside the tetragons), while the corresponding sectors in r$\gamma$GY show broader angles ($\sim114.854^\circ$) that are closer to the $120^\circ$ value associated with perfect $sp^2$ hybridization. Another important structural element is the bond angle at the atoms labeled 4 in Fig.~\ref{fig-01}f, that is, $\sim155.15^\circ$, which is about $25^\circ$ lower than the corresponding angle of the ideal $sp$ hybridization (as in the case of the atom labeled 3 in Fig.~\ref{fig-01}f). Because of this, the hybridization of atom 4 is intermediate between the $sp$ and $sp^2$ cases, although we will conventionally call it ``$sp$ atom'' since it has two first-nearest-neighbor ligand atoms. We point out, however, that this distortion relative to a truly $sp$ atom (such as atom 3 in Fig.~\ref{fig-01}f) is likely to affect the system's low electronic energy levels significantly. This will be verified when discussing the properties of the electronic structure. Finally, the other bond lengths of the system are $d_{13}\sim1.41$~\AA, $d_{24}\sim1.43$~\AA, $d_{33}\sim1.23$~\AA, and $d_{44}=1.24\sim$~\AA. We note that the bond length between a $sp$ and a $sp^2$ atom fluctuates over a narrow range, similar to the lengths of the $sp-sp$ bonds, which is a common property of other graphynic systems~\cite{oliveira2022}.

We also compare the stability of r$\gamma$GY with other systems in terms of formation energies. These quantities are calculated according to:
\begin{equation}
E^{\rm form}_1=\frac{E_{t}-N_{\rm atoms}\mu_C}{N_{\rm atoms}},
\end{equation}
where $E_t$ is the total energy of the system, $N_{\rm atoms}$ is the number of atoms per unit cell, and $\mu_C$ the energy per atom in graphene, considered carbon's most stable structure. In addition to r$\gamma$GY, we also calculated $E^{\rm form}_1$ for BPN, HGY, $\gamma$-GY, and a polyyne chain. The results are listed in order of increasing values in Table~\ref{tab1}. The values of $E^{\rm form}_1$ suggest that r$\gamma$GY is as stable as the previously synthesized HGY, and its stability is intermediate between that of BPN and of a polyyne chain (also observed experimentally). Compared with $\gamma$-GY, r$\gamma$GY is slightly less stable than its hexagonal counterpart.

\begin{table}[h!]
\caption{Formation energies per atom for graphene, BPN, $\gamma$-GY, r$\gamma$-GY, HGY, and a polyyne chain.}
\centering
\begin{tabular}{|c|c|c|c|}
\hline
System           & $E^{\rm form}_1$ (eV/atom) & System           & $E^{\rm form}_2$ (eV/atom)   \\
\hline                                 
Graphene         & 0.000             & Graphene         & 0.000   \\
\hline
BPN              & 0.483             & Polyyne          & 0.000   \\
\hline
$\gamma$-GY      & 0.752             & $\gamma$-GY      & 0.107   \\
\hline                                 
r$\gamma$GY      & 0.854             & r$\gamma$GY      & 0.209   \\
\hline                                             
HGY              & 0.854             & HGY              & 0.210  \\
\hline                                 
Polyyne          & 1.289             & BPN              & 0.483   \\
\hline                                 
\end{tabular}
\label{tab1}
\end{table}

Another way to compare the relative stability of these systems is by considering the different $sp$ and $sp^2$ compositions of these structures through different definitions for the chemical potentials of these atoms. This is done in terms of the synthesized systems composed purely of $sp$ or $sp^2$ hybridizations, namely, a polyyne chain and graphene, respectively. This alternative way of defining the formation energy is given by:
\begin{equation}
E^{\rm form}_2=\frac{E_{t}-N_{sp}\mu_{sp}-N_{sp^2}\mu_{sp^2}}{N_{\rm atoms}},
\end{equation}
where $\mu_{sp}$ ($\mu_{sp^2}$) is defined as the energy per atom in a polyyne chain (graphene sheet), and $N_{sp}/N_{sp^2}$ is the number of $sp/sp^2$ atoms, with $N_{\rm atoms}=N_{sp}+N_{sp^2}$. The results for $E^{\rm form}_2$ are also listed in order of increasing values for the different structures considered in Table~\ref{tab1}. This result also confirms that r$\gamma$GY and HGY have similar stability. However, the stability of these two systems is closer to that of the pair of reference systems (graphene and polyyne) than that of BPN. As we discussed earlier, this result is strongly influenced by the fact that the internal strain (induced by the presence of angles that are not strictly 120 degrees as in graphene) over the different ring types is better accommodated in the graphynic system as a result of the large pore structure.

We also studied the mechanical stability of r$\gamma$GY by computing the components of its elastic tensor. For such a computation, we introduce lattice distortions by applying uniaxial strain values of -0.01, -0.005, 0.000, 0.005, and 0.01 separately along the $x$ and $y$ directions over the plane. These strain values are defined as:
\begin{equation}
 \varepsilon_{xx}=\frac{l_x-l^0_x}{l^0_x},\quad\textrm{and}\quad \varepsilon_{yy}=\frac{l_y-l^0_y}{l^0_y},
\end{equation}
where $l^0_x$ and $l_x$ ($l^0_y$ and $l_y$) are the relaxed and strained lattice constants along the $x$ ($y$) direction, respectively, and $\varepsilon_{xx}$ ($\varepsilon_{yy}$) is the corresponding component of the strain tensor. We also consider the same values for the biaxial strain (where $\varepsilon_{xx}=\varepsilon_{yy}$). Since r$\gamma$GY is expected to be anisotropic, we also applied a shear stress ($\varepsilon_{xy}$). This is defined as the tangent of the variation of the angle (in radians) between the vectors defining the supercell (originally orthogonal in the relaxed system). For $\varepsilon_{xy}$, we used the same values considered for uniaxial and biaxial strains. For these systems, we calculated the elastic strain energy $U(\varepsilon)$ defined as the difference between the total energy of the strained and relaxed systems per unit area. For low-strain values, such as those used here, it is written as:
\begin{equation}
U(\varepsilon)=\frac{1}{2}C_{11}\varepsilon^2_{xx}+\frac{1}{2}C_{22}\varepsilon^2_{yy}+C_{12}\varepsilon_{xx}\varepsilon_{yy}+2C_{66}\varepsilon^2_{xy},
\end{equation}
where $C_{11}$, $C_{22}$, $C_{12}$, and $C_{66}$ are the components of the elastic tensor, where we used the Voigt notation ($1-xx$, $2-yy$, $6-xy$)~\cite{andrew2012}. A least-square fitting of the data from the strain simulations is used to determine the components of the elastic modulus tensor. For r$\gamma$GY, we obtain $C_{11}=243.19$ N/m, $C_{22}=139.62$ N/m, $C_{12}=22.82$ N/m, and $C_{66}=0.99$ N/m. To be mechanically stable, the components of the tensors of the elastic modulus must obey the Born-Huang criteria~\cite{born1954,ding2013,zhang2015}, namely $C_{11}C_{22}-C_{12}>0$ and $C_{66}>0$, which are verified for r$\gamma$GY.

\begin{figure}[ht!]
\includegraphics[width=\columnwidth]{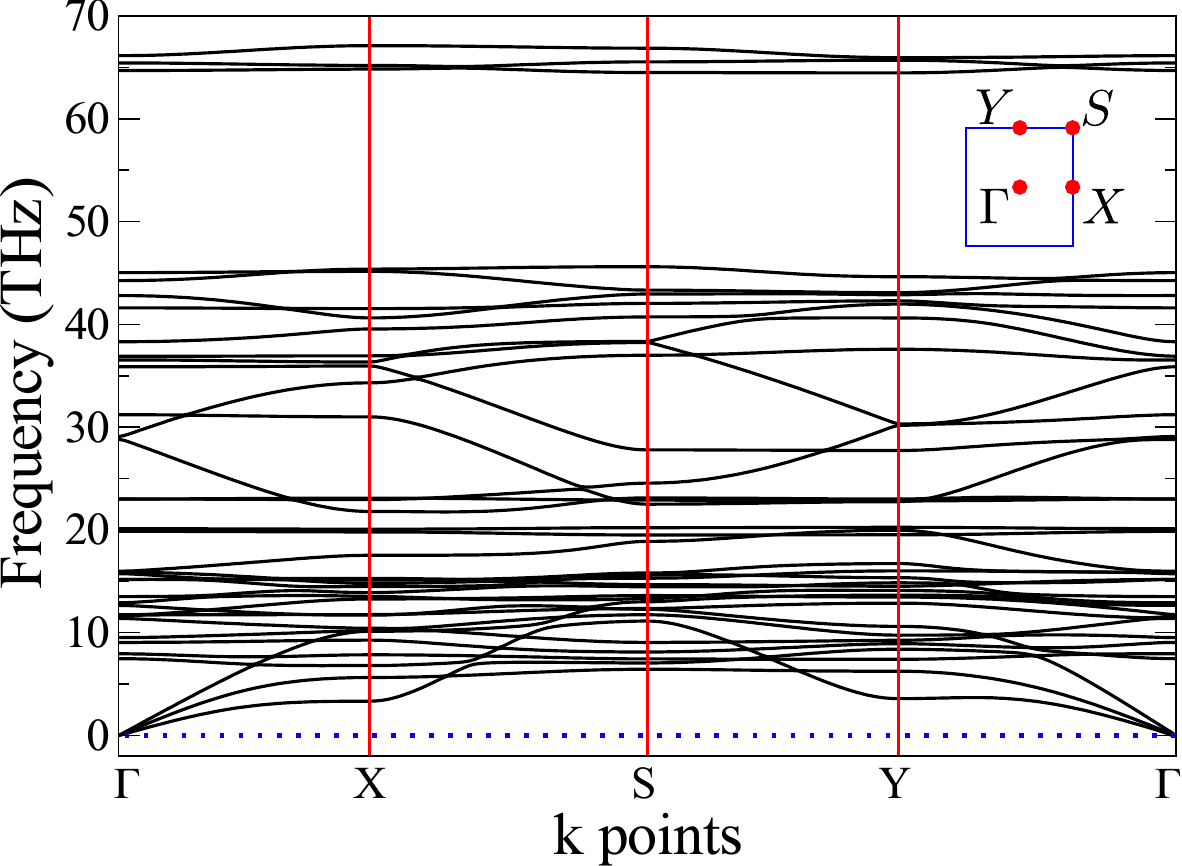}
\caption{Phonon band structure for the r$\gamma$GY sheet along the high-
symmetry lines of the Brillouin zone. The dynamical stability of the system is verified by the absence of complex frequencies (which would conventionally appear as negative frequencies in the graphical
representation). The acoustic branches along the $\Gamma-X$ and $\Gamma-Y$ feature a slightly different group velocity (see text). }
\label{fig-02}
\end{figure}

We also compute the phonon band structure for r$\gamma$GY. This is shown in Fig.~\ref{fig-02}, where we note the absence of imaginary modes, which would conventionally be shown as negative frequencies. This result indicates that r$\gamma$GY is expected to be dynamically stable. In addition, the phonons along the different directions clearly illustrate the anisotropic nature of the system. First, we note that as expected for a 2D system, one of the acoustic branch (ZA) is a flexural mode with a marked parabolic behavior. Second, we can can quantify the anisotropy by calculating the sound velocity (\textit{i.e.}, the slope of the linear acoustic branches) for the $\Gamma-X$ and $\Gamma-Y$ directions for the other two acoustic branches (A). The corresponding values for the $\Gamma-X$ direction are $\approx$ 2.213 km/s and 2.918 km/s, while the values for $\Gamma-Y$ are $\approx$ 2.110 km/s and 3.389 km/s. We note that the values for the second TA branch is about 16.1 \% higher for the $\Gamma-Y$ direction compared to $\Gamma-X$, respectively, while the velocity along $y$ for the second TA branch is approximately 4.7 \% lower than that along the $x$ direction, suggesting that thermal conductivities are anisotropic for r$\gamma$GY. As we will discuss later, the anisotropic character of r$\gamma$GY is even more apparent when considering electronic properties.


\begin{figure}[ht!]
\includegraphics[width=\columnwidth]{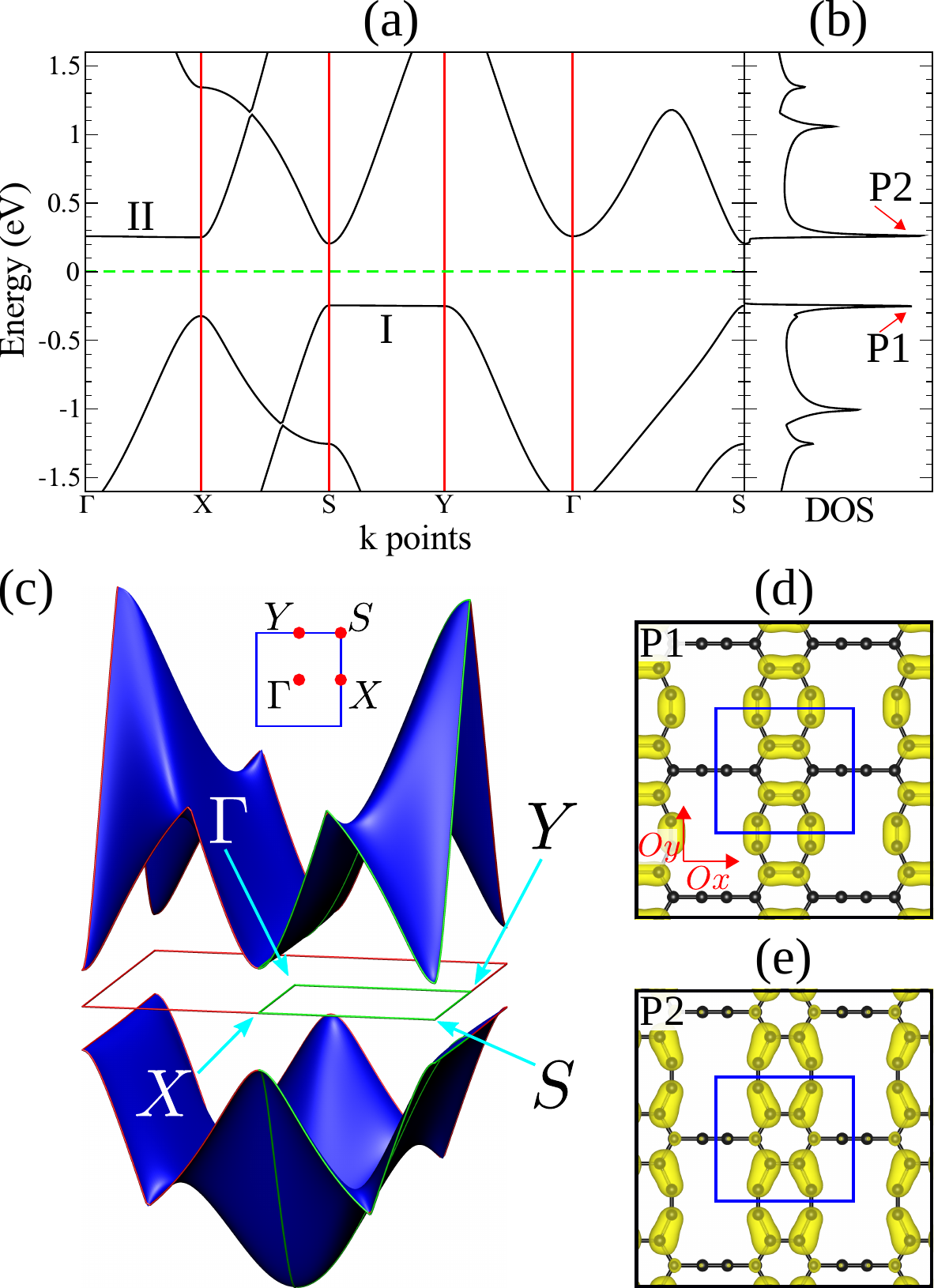}
\caption{(a) Electronic band structure of the r$\gamma$GY system along high-symmetry lines of the Brillouin zone and (b) the corresponding DOS as a function of energy. The P1 and P2 peaks in the DOS plot correspond to the flat bands along the $\Gamma-X$ and $S-Y$ paths. (c) Surface plot for the frontier bands of r$\gamma$GY over the entire Brillouin zone. The Fermi level and electronic bands along the edges of the Brillouin zone are highlighted by red lines, while the projections of the bands over the high-symmetry lines from the plot in (a) are represented by green lines. (d-e) LDOS plots for the energies of the P1 and P2 peaks.}
\label{fig-03}
\end{figure}

We now turn to the description of the electronic structure properties of r$\gamma$GY. The electronic band structure is shown in Fig.~\ref{fig-03}a, together with the corresponding density of states (DOS) shown in Fig.~\ref{fig-03}b, indicating that the system is a semiconductor with a 0.45 eV gap. The valence band maximum (VBM) corresponds to a flat state along the $S-Y$ path, which results in a van Hove singularity in the DOS, indicated by the P1 peak shown in Fig.~\ref{fig-03}b. The valence band also features a local maximum at the $X$ point of the BZ, which lies $\sim77$~meV below the flat band's global maximum. In addition, we note a flat region for the conduction band along the $\Gamma-X$ path. However, this flat region lies slightly above ($\sim35$~meV) the global minimum of the band, at the $S$ point. Note that the flat sector of the conduction band results in a prominent peak (P2) in the DOS plot in Fig.~\ref{fig-03}b and we note a small shoulder below it, which corresponds to the conduction band minimum (CBM) at $S$. These features can be seen more clearly in the plot reproduced in Fig.~\ref{fig-03}c, where we show the surface plots for the valence and conduction bands over the entire BZ. In Fig.~\ref{fig-03}d-e we also plot the local DOS (LDOS) for the P1 and P2 peaks, indicating that these states are located primarily around the tetragonal pores, but on complementary sets of bonds. Namely, P1 is located over the acetylenic bonds of the tetragonal pore and over the bonds shared by the tetragonal pores and the hexagons, while P2 lies over the bonds between the $sp$ and $sp^2$ atoms. We also note that these frontier flat states have negligible contributions from acetylenic bonds that link successive hexagons along the $a$ direction.  These P1 and P2 states are delocalized over quasi--1D sectors along the $b$ direction, showing negligible overlap along the orthogonal direction. Note that the acetylenic bonds along the $a$ direction are perfectly linear and are expected to constitute a more stable sector of the structure in comparison to the remaining $C\equiv C$ bonds of the system. This is because the acetylenic units that make up the tetragonal pores are not exactly linear. Therefore, this deviation from the ideal $sp$ geometry is expected to result in a stronger association of the frontier levels with these bonds than for the acetylenic links aligned with the direction $a$.

The set of quasi--1D states on this 2D structure is similar to those found in other systems~\cite{beserra2020,moreno2018}. Here, the lack of overlap for the electronic clouds along the $a$ direction is consistent with the flat feature of the frontier bands along the parallel $\Gamma-X$ and $S-X$ paths of the BZ (which are parallel to the reciprocal lattice vector corresponding to the $a$ direction in direct space).

\begin{figure}[ht!]
\includegraphics[width=\columnwidth]{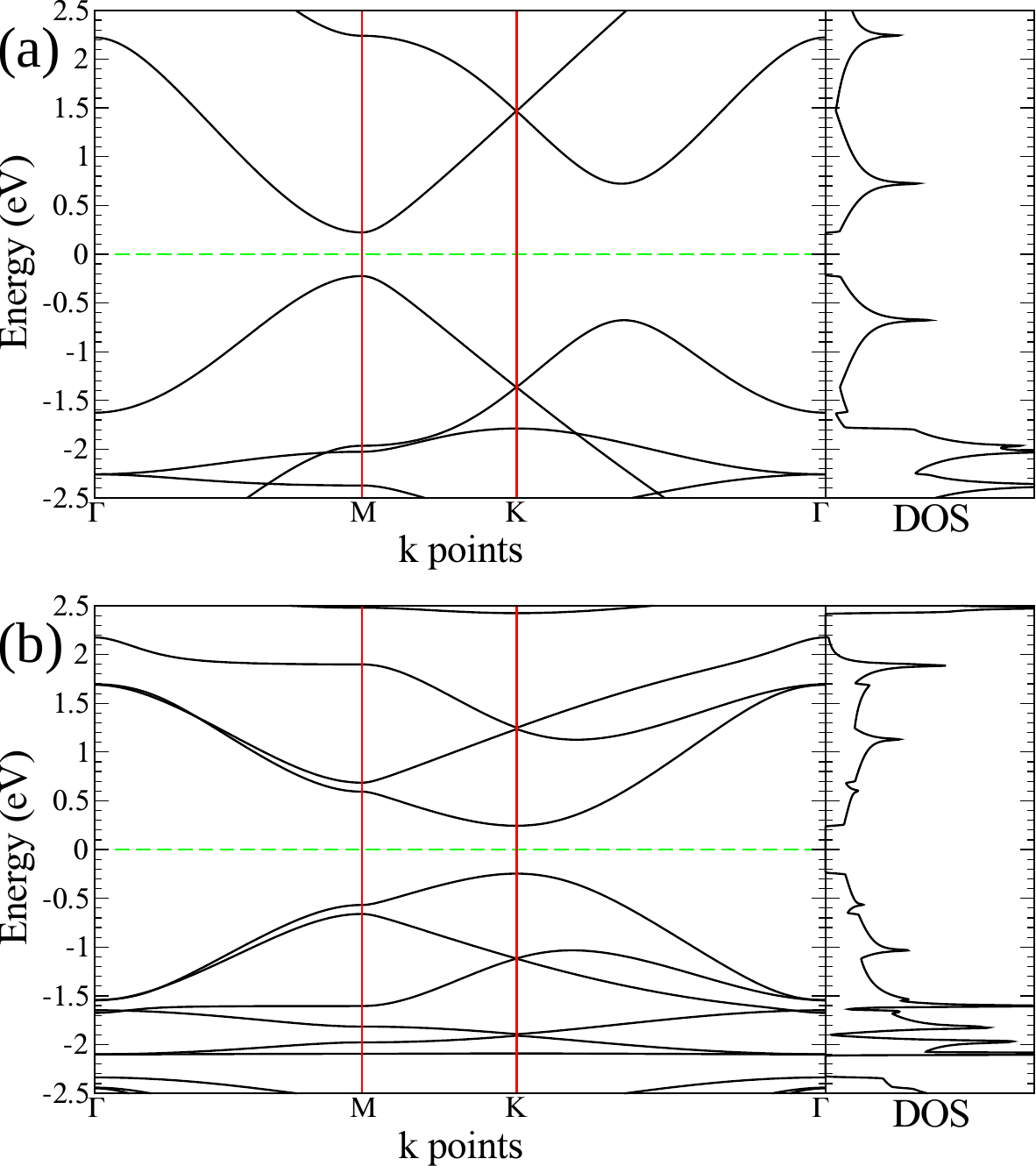}
\caption{(a) Electronic band structure along the high-symmetry lines of the Brillouin zone and DOS for the $\gamma$-GY sheet, and (b) for HGY.}
\label{fig-04}
\end{figure}

The semiconducting character of r$\gamma$GY is shared with its $\gamma$-GY and HGY counterparts, as shown in Fig.~\ref{fig-04}, where we plot the band structures of these two systems together with their corresponding DOS. The band gap for $\gamma$-GY (0.45 eV) is nearly the same as that of r$\gamma$GY, while the gap for HGY is slightly wider (0.49 eV). A marked difference between the $\gamma$-GY and HGY systems and the r$\gamma$GY sheet is that the latter is set on a rectangular lattice, while $\gamma$-GY and HGY are hexagonal structures. It turns out r$\gamma$GY features a highly anisotropic character for its frontier states, while the $E$ \emph{versus} $k$ relation features a more isotropic character for $\gamma$-GY and HGY. Furthermore, the frontier states for the rectangular system have a highly localized aspect, while the hexagonal systems are significantly more dispersive. In fact, the DOS features prominent peaks close to the system's VBM and CBM energies in r$\gamma$GY, while a step-like curve at the VBM-CBM is observed in the DOS of the hexagonal systems. The peaks that are closest to the Fermi energy in $\gamma$-GY and HGY are more than 0.5 eV away from $E_F$.

As it has already been shown for $\gamma$-GY, for example, the electronic properties of the GY structures can be tuned by changing the size of the acetylenic bridges that make up the backbone of the system. Therefore, we study the effect of introducing longer acetylenic chains (by adding another $-C\equiv C-$ unit) in the r$\gamma$GY lattice. We adopt three different strategies for adding a linking unit: 

\begin{enumerate}
 \item Only for the acetylenic links connecting successive hexagons along the $a$ direction (type-3 atoms);
 \item Only for the acetylenic links connecting successive hexagons along the $b$ direction (type-4 atoms); and
 \item For all the acetylenic links in the structure.
\end{enumerate}

\begin{figure}[ht!]
\includegraphics[width=\columnwidth]{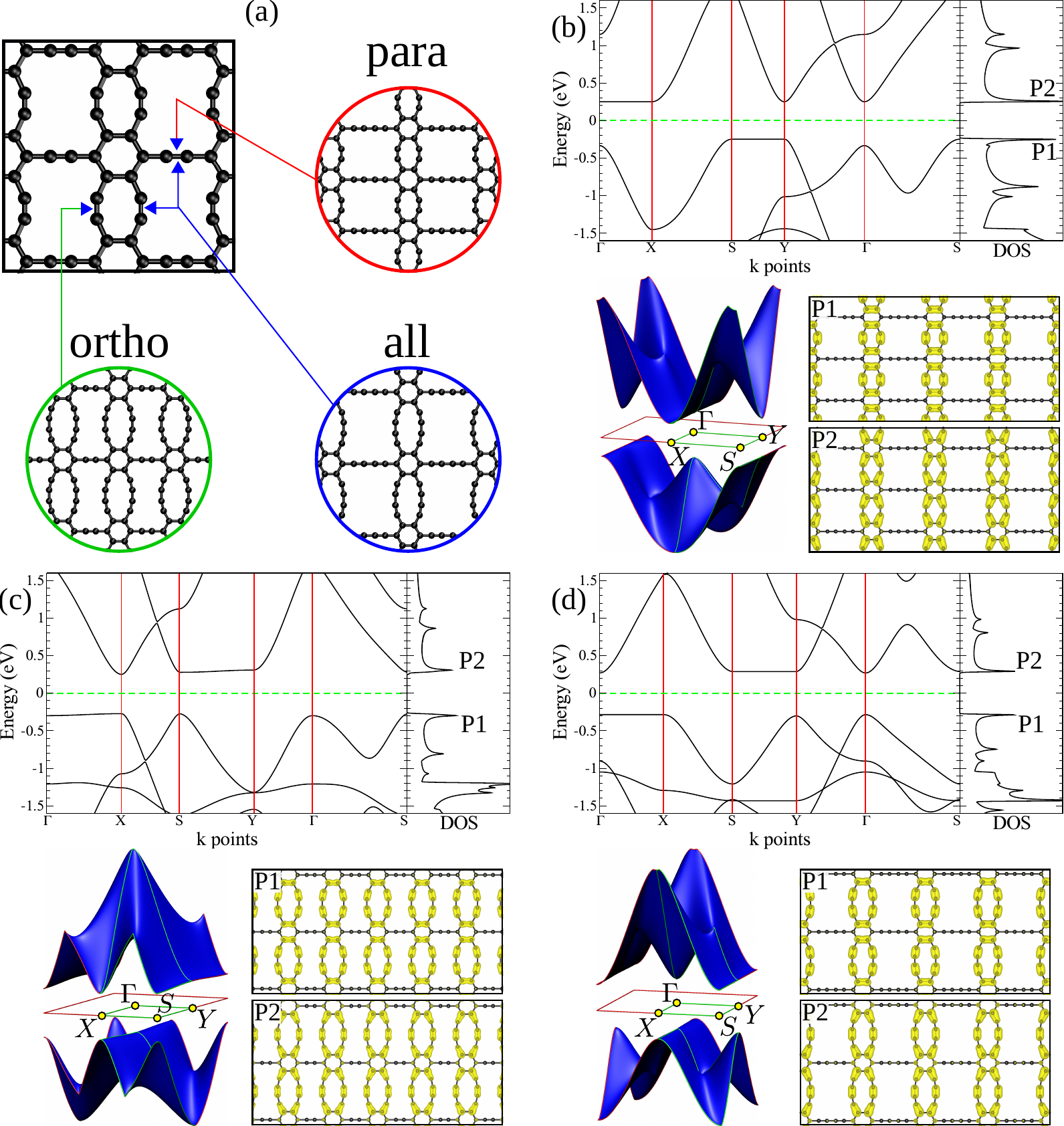}
\caption{(a) Schematic representation of how the $p$r$\gamma$GY, $o$r$\gamma$GY, and $a$r$\gamma$GY structures are defined from the parent r$\gamma$GY structure by the insertion of additional acetylenic units. (b) Electronic band structure along the high-symmetry lines of the BZ, DOS, surface plot for the frontier bands over the entire BZ, and LDOS plots for the P1 and P2 frontier states from the DOS for the $p$r$\gamma$GY system. (c) Same as (b) but for the $o$r$\gamma$GY sheet. (d) Same as (b) but for the $a$r$\gamma$GY system.}
\label{fig-05}
\end{figure} 

These structures are illustrated in Fig.~\ref{fig-05}a. The first two configurations are labeled $p$r$\gamma$GY and $o$r$\gamma$GY, where $p$ and $o$ stand for \emph{para} and \emph{ortho}, as inspired by the terminology for electrophilic aromatic substitution in benzene (considering the positions of the acetylenic bridges emerging from a hexagonal $sp^2$ ring). The third configuration will be referred to as $a$r$\gamma$GY, where $a$ stands for \emph{all} bonds emerging from the hexagonal rings. In Figs.~\ref{fig-05}b-d, we show the electronic band structures for these systems along the high-symmetry lines of the BZ, as well as the surface plots over the entire zone for the frontier bands. These three systems maintain the semiconducting character of the parent r$\gamma$GY structure, as well as the pair of flat bands along the $\Gamma-X$ and $S-Y$ directions. Each band structure plot in Fig.~\ref{fig-05} is also accompanied by the corresponding DOS plot, where the flat bands give rise to the emergence of two van Hove singularities similar to the P1 and P2 peaks described for r$\gamma$GY. The corresponding LDOS plots for these peaks are also illustrated in Fig.~\ref{fig-05}. The LDOS for these frontier levels is also spread like quasi-1D states along the $b$ direction of the sheets, similarly to the case of the r$\gamma$GY parent structure. These states are distributed over atoms located around the pores originating from the BPN tetragons, but the P1 and P2 states differ from each other by the bonds over which they overlap, in analogy to the parent r$\gamma$GY structure. 

The (direct) gaps for the $p$r$\gamma$GY, $o$r$\gamma$GY, and $a$r$\gamma$GY structures are 0.50~eV, 0.52~eV, and 0.55 eV, respectively. All these values are higher than the original r$\gamma$GY (0.45 eV). In particular, we note that the higher the ratio between the number of $sp$ and $sp^2$ atoms (\emph{all} $>$ \emph{ortho} $>$ \emph{para}), the wider the gap, an effect that is correlated with the greater degree of spatial isolation for the hexagonal $sp^2$ rings. In addition to the different gap values, the association of the flat band sectors along the $\Gamma-X$ and $S-Y$ paths to valence or conduction bands is different for the $o$r$\gamma$GY and $a$r$\gamma$GY systems compared to r$\gamma$GY, while this is not the case of the $p$r$\gamma$GY structure. 

\begin{figure*}[ht!]
\includegraphics[width=\textwidth]{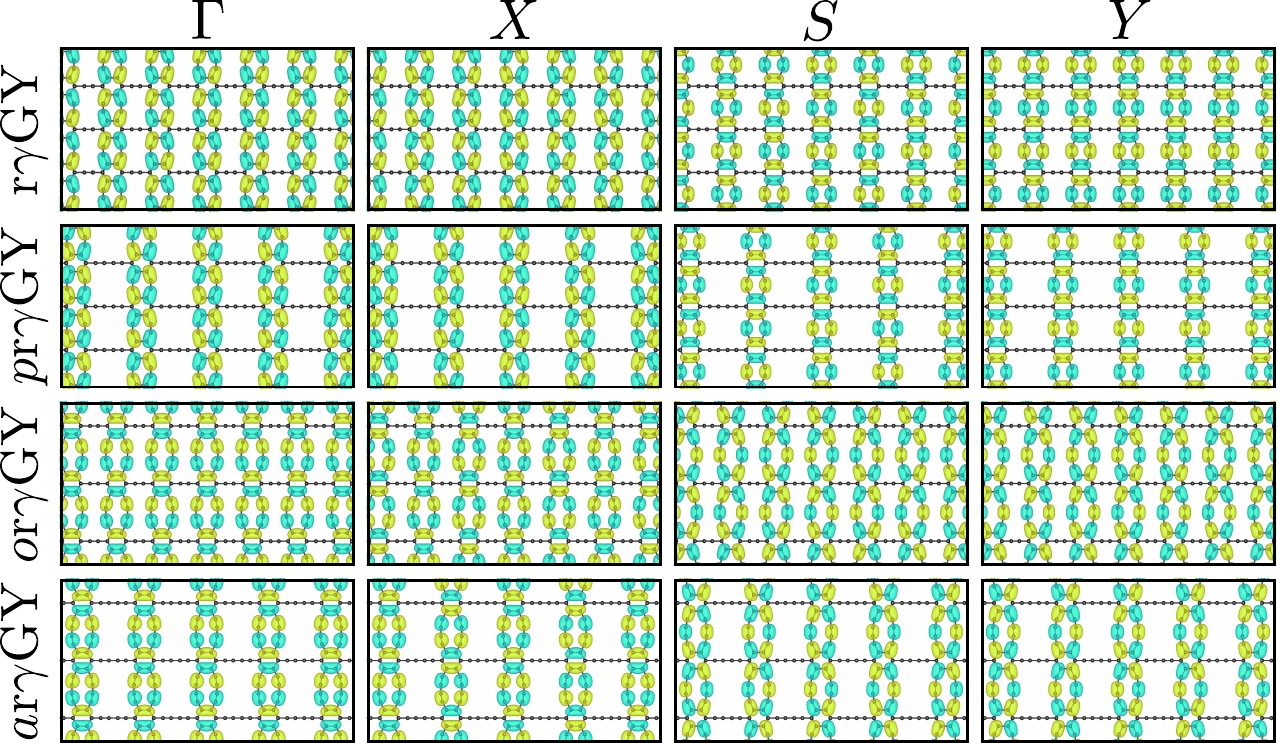}
\caption{Plots for the real part of the wavefunctions of the r$\gamma$GY, $p$r$\gamma$GY, $o$r$\gamma$GY, and $a$r$\gamma$GY structures for the flat frontier bands at the $\Gamma$, $X$, $S$, and $Y$ points of the Brillouin zone. Positive (negative) wavefunction values are represented by cyan (yellow) surfaces.}
\label{fig-06}
\end{figure*} 

To understand these inversions, we plot the real part of the wave functions for these bands at the $\Gamma$, $X$, $S$, and $Y$ points of the BZ for r$\gamma$GY and for the $p$r$\gamma$GY, $o$r$\gamma$GY, and $a$r$\gamma$GY structures in Fig.~\ref{fig-06}. First, we look at the states in r$\gamma$GY. The plots for this system in Fig.~\ref{fig-06} correspond to the flat part of the conduction (valence) band at $\Gamma$ and $X$ ($S$ and $Y$). In agreement with the LDOS plots shown in Fig.~\ref{fig-03}d-e, these wave functions are spread over the atoms located around the pores originating from BPN's tetragons. The lobes of the wave function of the conduction states lie on bonds involving one $sp^2$ atom of the hexagonal rings and one $sp$ atom of the acetylenic chain. As expected from the $e^{i\mathbf{k}\cdot\mathbf{R}}$ Bloch phase ($\mathbf{k}$ being a vector of the BZ and $\mathbf{R}$ a direct lattice vector), the lobes of the wave function at $\Gamma$ do not change sign when moving from one cell to another, while those of the wave function at $X$ (do not) invert their sign for neighboring unit cells along the $a$ ($b$) direction. The r$\gamma$GY valence states are seen to spread over the $sp^2-sp^2$ and $sp-sp$ bonds of the tetragonal pore. In addition, because of the Bloch phase, the lobes of the state $S$ change sign for neighboring cells along both $a$ and $b$ directions, while for the state at $Y$ this occurs only along the $b$ direction. 

Moving to the \emph{para} system, we observe that additional acetylenic links are added between the hexagons along the $a$ direction, where the wave functions corresponding to the flat bands have a negligible amplitude. As a result, this inclusion does not affect the symmetry of these states (as observed in Fig.~\ref{fig-06}) and the band structures of the r$\gamma$GY and $p$r$\gamma$GY states are qualitatively similar to each other. 

We also note that the lobes of the wave functions at $S$ and $Y$ over the two $sp^2-sp^2$ bonds inside a tetragonal pore of the r$\gamma$GY and $p$r$\gamma$GY systems carry the same sign, as they alternate with a lobe with the opposite sign over the $sp-sp$ bonds along the $b$ direction. In $o$r$\gamma$GY, we have two $-C\equiv C-$ bonds between the $sp^2-sp^2$ lobes, and the successive $-C\equiv C-$ lobes have opposite signs (since they show a nodal plane in the middle of the $\equiv C-C\equiv$ links). The results show that the two $sp^2-sp^2$ lobes inside a tetragonal pore of $o$r$\gamma$GY must have opposite signs (to accommodate successive nodal planes). However, these wave function configurations are no longer compatible with the symmetry of the $S$ and $Y$ points. Instead, they follow the symmetry at the $\Gamma$ and $X$ points. Similar arguments can be applied to the $\Gamma$ and $X$ states of r$\gamma$GY and $p$r$\gamma$GY, which become now compatible with the symmetry of the $S$ and $Y$ points in $o$r$\gamma$GY. As a result, the valence and conduction \textit{flat states} change their position over the BZ in the $o$r$\gamma$GY structure compared to the r$\gamma$GY and $p$r$\gamma$GY cases. Since the $a$r$\gamma$GY system can be obtained from the $o$r$\gamma$GY case by including additional acetylenic units between hexagons along the $a$ direction, this does not change the symmetry relations, and the $a$r$\gamma$GY and $o$r$\gamma$GY systems have similar band structures.

\section{Concluding remarks}
This study indicates that r$\gamma$GY is a nanostructure with the potential to be synthesized given its dynamical and mechanical stability. This carbon nanostructure features highly anisotropic properties, since its elastic properties, phonons, and the character of its frontier electronic states are remarkably different when considering different crystalline directions. In addition, r$\gamma$GY is a semiconducting system in its pristine 2D structure. This is different from many other 2D carbon allotropes that are usually metallic and require further chemical/physical modifications for a band gap to open. Finally, r$\gamma$GY features highly localized states whose symmetry can be further tuned by changing the size of the acetylenic chains and by the strategic choice of their location.

\section{Acknowledgments}
E.C.G acknowledges the support of the Brazilian agency CNPq (Process No. 310394/2020-1). ECG acknowledges the support from Funda\c c\~ao de Amparo \`a Pesquisa do Estado do Piau'i (FAPEPI) and CNPq through the PRONEM program. The authors thank the Laborat\   'orio de Simula\c c\~ao Computacional Caju\'ina (LSCC) at Universidade Federal do Piau\'i for computational support. The authors also acknowledge computational support from Centro Nacional de Processamento de Alto Desempenho at Ceará (CENAPAD-UFC) and São Paulo (CENAPAD-SP). 

\bibliographystyle{edu_style2}
\bibliography{bibs}

\end{document}